# Strain-gradient-induced magnetic anisotropy in straight-stripe mixed-phase bismuth ferrites: An insight into flexomagnetic phenomenon

Jin Hong Lee,[1,*] Kwang-Eun Kim,[1] Byung-Kweon Jang,[1] Ahmet A. Ünal,[2] Sergio Valencia,[2] Florian Kronast,[2] Kyung-Tae Ko,[3] Stefan Kowarik,[4] Jan Seidel,[5] and Chan-Ho Yang[1,6,†]

[1] *Department of Physics, KAIST, Yuseong-gu, Daejeon 34141, Republic of Korea*
[2] *Helmholtz Zentrum Berlin für Materialien und Energie, Albert-Einstein-Strasse 15, D-12489 Berlin, Germany*
[3] *Max Plank POSTECH Center for Complex Phase Materials & Department of Physics, POSTECH, Pohang, Gyeongbuk 37673, Republic of Korea*
[4] *Institut für Physik, Humboldt-Universität zu Berlin, Newtonstrasse 15, D-12489 Berlin, Germany*
[5] *School of Materials Science and Engineering, University of New South Wales, Sydney, New South Wales 2052, Australia*
[6] *KAIST Institute for the NanoCentury, KAIST, Yuseong-gu, Daejeon 34141, Republic of Korea*

*E-mail: rifle@kaist.ac.kr
†E-mail: chyang@kaist.ac.kr

# ABSTRACT

Implementation of antiferromagnetic compounds as active elements in spintronics has been hindered by their insensitive nature against external perturbations which causes difficulties in switching among different antiferromagnetic spin configurations. Electrically-controllable strain gradient can become a key parameter to tune the antiferromagnetic states of multiferroic materials. We have discovered a correlation between an electrically-written straight-stripe mixed-phase boundary and an in-plane antiferromagnetic spin axis in highly-elongated La-5%-doped $BiFeO_3$ thin films by performing polarization-dependent photoemission electron microscopy in conjunction with cluster model calculations. Model Hamiltonian calculation for the single-ion anisotropy including the spin-orbit interaction has been performed to figure out the physical origin of the link between the strain gradient present in the mixed phase area and its antiferromagnetic spin axis. Our findings enable estimation of the strain-gradient-induced magnetic anisotropy energy per Fe ion at around $5\times10^{-12}$ eV m, and provide a new pathway towards an electric-field-induced 90° rotation of antiferromagnetic spin axis at room temperature by flexomagnetism.

# I. INTRODUCTION

Although the majority of magnetic ordering found in solid state systems is antiferromagnetic (AFM) [1], the intrinsic absence of a measurable net magnetization has caused difficulties both in detecting and hence controlling AFM properties [2, 3]. During the last decade, significant efforts have been made to exploit the rigidity of antiferromagnets with respect to external perturbations for technological uses, and consequently antiferromagnets have been employed as a *static* part in multilayered systems such as spin valves [4], magnetic tunnel junctions [5] and spin-orbit torque devices [6]. Under these circumstances, the precise control and switching among different AFM configurations are very useful for future spintronic applications. Some recent works have demonstrated the writing and reading processes between two AFM configurations can be driven by short femtosecond laser pulses and time-resolved linear magnetic birefringence [7], stochastic spin-polarized tunneling current and spin-polarized scanning tunneling microscope imaging [8], AFM-to-ferromagnetic transition and anisotropic magnetoresistance [9], and staggered current-induced field and anisotropic magnetoresistance [10].

In this context, the strain-driven morphotropic phase boundary (MPB, also known as mixed phase areas) found in the multiferroic $BiFeO_3$ (BFO) [11–13] is of importance. It has potential merits of low switching power consumption as an AFM insulator as well as the longer retention time compared to the multiferroic switching via non-180° polar rotation which creates a significant ferroelastic strain with respect to surrounding regions [14, 15]. Furthermore, it has been found that straight-stripe MPBs of two constituent multiferroic phases can be created by an external electric field to be over tens of microns along specific in-plane directions with a uniform boundary density in La-5%-doped BFO (BLFO) [16]. This deterministic control not only improves the feasibility of magnetoelectric and piezomagnetic applications but also provokes discussion on flexomagnetism associated with the controllable interfacial strain gradient as much as $\sim 3 \times 10^7$ m$^{-1}$ [17] which is the largest value observed in rigid dielectric materials. To proceed toward the reliable control of AFM configurations, it is required to unravel the magnetic state inherent in the straight-stripe mixed-phase areas and its correlation with the strain gradient. Despite its importance, the direct observation of the AFM spin axis within the nanoscale structure remains challenging because of its complex crystalline structure and small size restricting conventional magnetic characterization techniques.

In this article, we report that electrical switching of MPBs can be accompanied by a 90° rotation of their AFM spin axis. The in-plane AFM spin axis of the written MPB area is determined to be parallel to either [100] or [010] crystalline axis depending on the MPB elongation axis, offering a useful pathway toward the electric switching of the in-plane AFM spin axis. A single-ion anisotropy calculation shows that the magnetic anisotropy can be induced by the strain-gradient-induced Fe off-center displacement at MPB and explains how the AFM spin axes in MPB regions become almost perpendicular to the MPB elongation axes as a consequence of the flexomagnetic coupling.

BFO, which has a rhombohedral structure ($R3c$) and displays simultaneous ferroelectric ($P \sim 90$ μC cm$^{-2}$, $T_C \sim 1100$ K) and AFM ordering (G-type, $T_N \sim 643$ K), is one of few room-temperature multiferroic compounds [18–21]. Since BFO provides a strongly-coupled charge-spin-lattice system at room temperature, its polymorphic phases, e.g., a highly-elongated tetragonal-like metastable phase (T-phase, $T_N \sim 380$ K) [22–26], a rhombohedral-like monoclinic phase [27–30] and an orthorhombic phase [31–33], have been intensively investigated in the exploration of multifunctional phenomena. Especially, Zeches *et al*. have reported a striking observation in compressive-strained BFO thin films deposited on (001)-oriented LaAlO$_3$ (LAO) substrates: a defect-less boundary between T-phase ($c/a \sim$ 1.23) and another polymorphic phase ($c/a \sim$ 1.07, S-phase; here, we follow the notation in Ref. 34) where a large strain gradient appears [11]. The boundary is what we call MPB, which has brought intriguing phenomena such as colossal electromechanical response [35, 36], occurrence of spontaneous magnetic moment [37], enhancement of electronic conduction [16, 38], strain-gradient-induced anisotropic photocurrent [17], and reversible electrochromism [39]. By performing a systematic angle-resolved piezoresponse force microscopy (PFM) study in BLFO thin films on LAO substrates, Kim *et al.* argued that a slow-scan direction during PFM-based electrical poling determines the major elongation axis of MPBs and it is possible to create uniform long-and-straight stripes [16]. By virtue of the excellent MPB controllability of BLFO thin films, we have been able to survey a correlation between the MPB elongation axis and its in-plane AFM spin axis.

## II. EXPERIMENTAL METHODS

A 40-nm-thick BLFO thin film buffered with a 5-nm-thick Pr$_{0.5}$Ca$_{0.5}$MnO$_3$ bottom electrode was deposited on a (001)-oriented LAO substrate by using pulsed laser deposition with a 248-nm KrF excimer laser. Spatially-resolved x-ray absorption spectroscopy (XAS) and x-ray linear dichroism (XLD) measurements, using linearly-polarized photons in the region of Fe *L*-edges, were done by means of photoemission electron microscopy (PEEM) at the UE-49 PGMa beamline of the BESSY II storage ring of the Helmholtz-Zentrum Berlin [40, 41]. At room temperature, the incident photons with $\sigma$ polarization ($\vec{E} \parallel ab$-plane) impinged the sample surface at a polar angle of 74° from the film normal. XAS and XLD data were obtained as function of the relative angle between the electric-field axis of the radiation ($\vec{E}$) and the MPB elongation axes. For full-multiplet configuration-interaction cluster model calculations, we employed XTLS9.0 with a FeO$_5$ crystal field (square pyramidal site symmetry of T-phase BFO; $C_{4v}$) [42, 43]. A T-phase model cage with (1) a shear distortion along the crystalline $a$ axis (M$_C$) by a monoclinic angle of 2° and (2) an exchange field of 12 meV along the crystalline $b$ axis was used in the calculations; in other words, our model has a monoclinic tilt along the crystalline $a$ axis and its AFM spin axis $\vec{S}$ is parallel to the crystalline $b$ axis.

## III. RESULTS

### A. Electrical aligning of four MPB elongation axes

In Fig. 1(a), a reciprocal space map around the (002) diffraction position of the as-grown BLFO film shows multiple peaks originating from the T-phase matrix (labeled as T$_0$) and the mixed-phase regions (labeled as T and S). The four S peaks at $L$ = 1.81 infer the existence of four MPB elongation axes, and other directions of MPB elongation axes are not allowed in the film; for detailed information of the reciprocal space map analysis, we refer to previous works [44, 45]. A uniformly-aligned large area including all possible four MPB elongation axes was required to carry out a systematic XLD analysis on each MPB elongation axis. For the purpose of aligning MPBs, we performed a 30°-rotated double-box poling with a −10-V-biased conducting tip along two orthogonal slow-scan directions (blue-colored hollow arrows) and measured an out-of-plane PFM image after the electrical poling [Fig. 1(b)]; we chose 30° by inspecting the relation between the slow-scan direction and the MPB alignment. A surface image of the double-box-poled area and its surrounding shows two major features: four randomly-distributed MPB elongation axes in the as-grown region and four uniformly-aligned MPB elongation axes within the double-box-poled area [Fig. 1(c)]. Here, we classify the electrically-written MPB area by four polygons with different colors, and each color bar inside the polygons represents the elongation axis of MPBs. A schematic summary of the MPB elongation axes is described on the top-right corner, and each MPB elongation axis has a 12°-offset from [100] or [010] crystalline axis.

## B. XMLD-PEEM investigation on the electrically-written MPB regions

For the local XAS experiments, the linearly-polarized x-ray beam illuminated the whole PFM-written double-box area and hence all the four MPB regions could be probed simultaneously in the same field of view. The electric-field axis ($\vec{E}$) of the photons lays within the sample surface, and spatially-resolved XAS data were obtained at four different sample azimuthal angles: (1) $\vec{E}$ ∥ the green-colored MPB, (2) $\vec{E}$ ∥ the black-colored MPB, (3) $\vec{E}$ ∥ [010], and (4) $\vec{E}$ ∥ [100]. Figure 2(a) depicts the four local Fe $L$-edge XAS spectra averaged from each MPB region when $\vec{E}$ is parallel to the green-colored MPB axis. The XAS spectra show two main spectral features related to the $L_3$ and $L_2$ edges (corresponding to $2p_{3/2}$ → $3d$ and $2p_{1/2}$ → $3d$ dipole transitions) which stem from the $2p$ core-hole spin-orbit splitting. On the right side, a zoomed-in graph of the $L_2$ edge is displayed and we observe two subpeaks ("white lines") originating from the final state multiplets of $Fe^{3+}$ ($3d^5$) under the crystal field. Interestingly, the $L_2$-edge spectral shape depends on whether it corresponds to the inner-box or outer-box regions of the PFM-written double box. The $L_2$-edge XAS signals corresponding to the inner-box region (black-solid and red-dashed curves almost parallel to [100]) display a higher intensity at 720.55 eV (subpeak A) but a lower intensity at 722.2 eV (subpeak B) as compared with the XAS signals of the outer-box region (green-solid and blue-dashed curves almost parallel to [010]). Since the XAS signal difference is very subtle, we display XLD-PEEM images at the aforementioned four different $\vec{E}$-to-sample geometries in Fig. 2(b); the image contrast is defined as $I_{XLD-PEEM} = (I_B − I_A)/(I_B + I_A)$, and it reflects the local XLD signals [41]. In every XLD-PEEM image, the contrast between the inner-box region and the outer-box region of the double box clearly appears so that we can readily match the local positions with the surface image in Fig. 1(c). Upon closer inspections of the four XLD-PEEM images, we observe that despite a discernable interior angle 24° between their respective MPB elongation axes, the contrast of black and red MPB regions (the inner-box region) is always equal. If we recall the relation $I_{\text{XLD-PEEM}} \propto \cos^2\theta$ [46], the contrast could vary by 17% when $\overleftrightarrow{E}$ is parallel to one of the MPB elongation axes. The same is true for green and blue MPB regions of the outer-box region.

Here we should consider that the measured XLD-PEEM contrast can originate either from an in-plane structural anisotropy of the MPB regions (x-ray natural linear dichroism; XNLD) or an aligned in-plane AFM spin axis (x-ray magnetic linear dichroism; XMLD), or from both. If the XNLD contribution from the MPB regions was significant, the observed XLD-PEEM contrasts from the double-box region would not be grouped into two as we observed, rather we would observe four contrasts from the four different MPB elongation axes. The presence of only two different contrast levels provides us strong evidence that the larger contribution to the observed XLD-PEEM contrast arises from the aligned in-plane AFM spin axis (XMLD) within the MPB regions. Another feature from the XLD-PEEM images is that the contrasts of the double-box region become stronger in the cases when $\overleftrightarrow{E}$ ∥ [100] and $\overleftrightarrow{E}$ ∥ [010] than in the other two cases when $\overleftrightarrow{E}$ is 12° off from either [100] or [010]. This would suggest that each MPB region generate an in-plane AFM spin anisotropy along the [100] crystalline axis (or [010], depending on the MPB elongation axis) but not along any other intermediate directions, and one MPB region strongly correlates with the aligning axis of AFM spins.

## C.   In-plane AFM spin axes of the electrically-written MPB regions

In order to determine the AFM spin axis of each MPB region, cluster model calculations based on a T-phase unit cell were employed and compared with the experimental data. We chose a T-phase model on behalf of the unit cells in MPB regions since the electrically-written MPB regions are mainly composed of T and S-phase (~93% of the whole area), and the in-plane *a/b*-axis anisotropy of the two phases is comparable [44, 47]. Figures 3(a) and (b) show the calculation results of XAS, XNLD and XNLD+XMLD. We observe that the calculated XNLD signals are about two orders of magnitude smaller than the calculated XMLD signals. Also, the sign of XMLD obeys the geometric XMLD rule of iron oxides [48, 49].

Figures 3(c) and (d) display our experimental XLD spectra corresponding to configurations with $\vec{E} \parallel [100]$ and $\vec{E} \parallel [010]$, respectively. Here, the XLD spectra display the difference between the XAS intensities of two perpendicular MPB groups (the inner-box region and the outer-box region). The four experimental XLD spectra under one electric-field axis give the identical line shape, which is an expected result from the above XLD-PEEM analysis. For a 90° rotation of the incident $\sigma$ polarization, the XLD signals reverse their sign. A comparison between Figs. 3(b) and (d) reveals that the $L_2$-edge XLD line shapes are very similar and the $L_2$-edge XLD amplitude is ~65% of the calculated XMLD so we can reasonably extract the in-plane AFM spin axis of the MPB region. We note that although the experimental $L_3$-edge XLD signals at 707.7 eV and 709.2 eV agree with the calculation on the whole, a reliable analysis on the $L_3$-edge signals was relatively challenging because the sharper and stronger line shape of the dichroism was sensitive to secondary multiplet features and vulnerable to photon energy misalignment. Figure 3(e) summarizes our result that the black and the red horizontal MPB regions (the inner-box region) share a [010] AFM spin axis, and the green and the blue vertical MPB regions (the outer-box region) share an [100] AFM spin axis.

### D. Magnetic anisotropy induced by the flexoelectric cation shift at MPB

From our observation, one specific question arises: what is the microscopic origin of the almost perpendicular relation between the MPB elongation axis and its AFM spin axis? Chu *et al*. [17] discussed that the giant strain gradient (~$3\times10^7$ m$^{-1}$) existing in MPB enforces Fe ions to shift perpendicular to the MPB elongation axis and to hybridize strongly with the nearest oxygen in the oxygen cage, thereby creating a large anisotropic photocurrent at MPB. Accordingly, the in-plane projection of Fe off-centering axis (with respect to the oxygen cage) in the inner-box region should be perpendicular to that in the outer-box region. If the Fe off-centering of one MPB strongly constrains its AFM spin axis to lie parallel to the off-centering direction, spins in the surrounding T and S-phase regions will be locked toward the same AFM axis of MPB via the exchange interaction. At this moment, it is noteworthy mentioning that our observation does not simply remain the determination of the spin axis of a specific material, but provides a general insight into the enhancement of the spin-orbit-interaction-based effects such as the Dzyaloshinskii-Moriya interaction [50], the Rashba effect [51] as well as the single-ion magnetic anisotropy when the magnetic ions of interest are located at an asymmetric position by significantly evading the angular momentum quenching in solids.

In order to examine the underlying physics by explicitly capturing the role of the asymmetric environment around the transition metal ion, a single-ion anisotropy model calculation was conducted as a function of Fe off-center displacement. We employ a one-electron Hamiltonian for the Fe 3*d* levels including the spin-orbit interaction:

$$H = V_{\text{CF}} - J\vec{S}\cdot\hat{\alpha} + \xi\vec{L}\cdot\vec{S}, \tag{1}$$

where the three terms correspond to the crystal field potential, the effective exchange energy, and the spin-orbit coupling energy ($\xi = 0.05$ eV [52]). In the second term, the molecular exchange field $J\hat{\alpha}$ represents the net effect from the AFM spin ordering of all other Fe ions and the strength $J$ is given by a large enough value, *e.g.*, 4 eV, to ensure that the resultant magnetization is toward the given control direction of $\hat{\alpha}$. In reality, the effective exchange strength approximately corresponds to the exchange coupling constant (a few meV as we used in the cluster model calculation before) times the coordination number in the mean-field regime, however, it is still valid because the total energy offset arising from the arbitrarily large choice is completely neglected at the moment that two independent total energies calculated at different directions are subtracted from each other to get the anisotropy energy. By comparing total energies ($E_{\text{tot}}^{\alpha}$) calculated under three different AFM spin axes, *i.e.*, $\hat{\alpha} = \{\hat{x}, \hat{y}, \hat{z}\}$, we estimate the magnetic anisotropy energy of a given unit cell. The details of our calculation are provided in Appendix A.

As shown in Fig. 4(a), the starting structural model is a tetragonal structure with a square pyramidal ($C_{4v}$) site symmetry. In order to construct a simple model for a unit cell located at MPB, we set $a = b = 3.8$ Å and $c = 4.4$ Å which are the average values of the lattice parameters of T and S-phase unit cells [11]. The O-cage and Bi-cage centers are shifted by 0.35 Å and −0.15 Å along the $z$ axis from the origin at the Fe position; these values are about half of the corresponding T-phase data [53]. Then, an in-plane Fe off-centering along the $y$ axis is introduced to investigate the change of the single-ion anisotropy; in order to fix the origin at the Fe position, we relatively shift the Bi and O cages by $\delta$ to the opposite direction of the Fe off-centering. The additional electrons transferred from O 2$p$ to Fe 3$d$ due to the strong Fe–O hybridization are also considered so that we can adjust the oxidation state of Fe at MPB to +2.65 as previously reported [54]. Figure 4(b) shows the magnetic anisotropy energy between the $\hat{\alpha} = \hat{y}$ case and the other two cases. The result displays that the total energy has the lowest value when the spin axis is parallel to the off-centering axis.

To further understand the role of the spin orbit coupling, we investigate the $\delta$-dependence of the orbital angular momentum and spin, *i.e.*, $\langle L_x \rangle$, $\langle L_y \rangle$, $\langle L_z \rangle$ and $\langle S_\alpha \rangle$, under $\hat{\alpha} = \{\hat{x}, \hat{y}, \hat{z}\}$ cases. Figures 5(a-c) show that the orbital angular momentum components do not monotonically change with $\delta$, and they strongly depend on both the site symmetry of a transition metal ion and the distances between the transition metal ion and surrounding oxygen ions. Since the Fe–O hybridization effect becomes stronger for the larger Fe off-centering, the magnitude of the total spin deviates more from 2.5 (5 electrons × $\hbar/2$) by transferring more down spins from the surrounding oxygen ions (Fig. 5d). We note that the magnitude of the total orbital angular momentum is not conserved when the spin direction is rotated, indicating the spatial distributions of electron clouds are influenced by the spin direction through the spin-orbit interaction. This modification affects not only the spin-orbit coupling energy but also the crystal field potential energy. Therefore, the role of the spin-orbit interaction in determining the magnetic anisotropy energy of a system becomes more significant when the Fe off-centering becomes larger, which helps the system to decrease the total energy when the spins are parallel to the off-centering axis (or the strongest bonding axis).

If we assume that the previously estimated flexoelectric polarization ~30 μC cm$^{-2}$ [17] at MPB is driven solely by the Fe off-centering, we obtain $\delta \sim 0.4$ Å and the corresponding magnetic anisotropy energy between the in-plane *x* and *y* axes becomes ~140 μeV/Fe. In order to get a sense of the magnitude of the calculated magnetic anisotropy energy, it is noteworthy to mention that the magnetic anisotropy energy of MPB is larger than that of a magnetite but smaller than that of a hematite; room-temperature magnetocrystalline anisotropy constants of a magnetite and a hematite are ~14 µeV/Fe and ~1.1 meV/Fe, respectively [55]. The low symmetry of the cation sites in hematite allows spin-orbit coupling to cause the abnormally large magnetic anisotropy as well as canting of moments above the Morin transition. We also estimate a staggered coercive field ($H_{\text{stag}}$) to rotate the spins from $y$ axis (easy axis) to $x$ axis (hard axis) by application of a staggered magnetic field, *i.e.*, a virtual anti-parallel field commensurate with the antiferromagnetic order. For $\delta \sim 0.4$ Å, the total spin along $y$ axis, *i.e.*, $\langle S_y \rangle$, is calculated to be 2.32 per Fe ion, and this corresponds to 4.64 $\mu_B$/Fe so we obtain 0.52 T for the staggered coercive field (4.64 $\mu_B$/Fe × $H_{\text{stag}}$ = 140 µeV/Fe). Considering the strain gradient in the system is ~$3 \times 10^7$ m$^{-1}$, we can lead to the conclusion that the strain-gradient-induced magnetic anisotropy energy per Fe ion is of the order of magnitude of $5 \times 10^{-12}$ eV m.

Based on our calculation, we propose that the magnetic anisotropy from the periodic structure of MPBs makes the surrounding AFM spins of T and S-phase be almost perpendicular to the MPB elongation axis and simultaneously be parallel to the crystallographic axis ([100] or [010]) [Fig. 4(c)]. Our observation offers a new insight into the enhancement of the spin-orbit interaction effect by strain gradients. The main finding is that if the transition metal ion is located at an asymmetric position within the octahedron, the spin-orbit interaction effects including the single-ion magnetic anisotropy become more significant because of a non-quenched orbital angular momentum. There are various ways to induce the asymmetric environment in addition to the strain-gradient-induced cation off-centering. External or built-in electric fields in asymmetric device structures and intrinsic electric fields in inversion-symmetry-broken materials can induce such asymmetric positioning which results in considerable Rashba effects, and oxygen buckling also plays a similar role in the emergence of a significant asymmetric exchange interaction. The strain gradient is another control parameter to break the symmetry, thereby enhancing the spin-orbit effect without changing the constituent element to a heavier one in order to increase the spin-orbit coupling strength. In addition, previous reports which demonstrate the mechanical switching of MPB elongation axes [56, 57] hold promise for new mechanomagnetic phenomena such as rotating the AFM spin axis by applying a local pressure or bending the film [58].

## IV. CONCLUSIONS

We found that the electric-field-induced alignment of the MPB elongation axes accompanies the alignment of AFM spin axes, and a reversible 90° rotation is successfully demonstrated in straight-stripe mixed-phase BLFO thin films. The local AFM spin axis measurement by using PEEM enabled observing the subtle XLD signals from the aligned MPB elongation axes. We could determine the parallel relation between the strain-gradient-induced Fe off-centering and the AFM spin axis by employing both cluster model and single-ion magnetic anisotropy calculations. Our discovery opens the door into the flexomagnetism connecting magnetism with strain gradients.

## ACKNOWLEDGMENTS

This work was supported by the National Research Foundation (NRF) of Korea funded by the Korean Government (Contracts Nos. NRF-2014R1A2A2A01005979 and NRF-2013S1A2A2035418). This work was also supported by the NRF via the Center for Quantum Coherence in Condensed Matter (2016R1A5A1008184). J.S. acknowledges funding from the Australian Research Council through Discovery grants and a Future Fellowship. The Helmholtz-Zentrum Berlin, including BESSY II, is supported by the BMBF.

# APPENDIX A: TRANSITION-METAL-ION OFF-CENTER EFFECT ON THE SINGLE-ION ANISOTROPY OF BFO

In order to understand our observation of the almost perpendicular relation between the MPB elongation axis and the spin easy axis, we performed a theoretical calculation on in-plane magnetic anisotropy associated with the flexoelectric Fe off-centering (with respect to the oxygen cage) at MPBs. The large discrepancy between the *c*-axis lattice parameters of T and S-phase generates the colossal strain gradient at the polymorphic interface and the Fe off-centering occurs perpendicular to the MPB elongation axis [17]. Here, we investigate how the Fe off-centering can influence on in-plane magnetic anisotropy as a result of the enhancement of the spin-orbit coupling effect.

To construct the one-electron Hamiltonian for Fe 3*d* levels, we consider three terms, *i.e.*, the crystal field potential from Bi and O ions surrounding a Fe ion of interest (including the Fe off-centering effect), the effective exchange interaction with surrounding Fe ions, and the spin-orbit coupling that gives a chance to the inter-mixing between spin-up and spin-down states:

$$H = V_{\mathrm{CF}} - J\vec{S}\cdot\hat{\alpha} + \xi\vec{L}\cdot\vec{S}. \tag{A1}$$

The parameters in the Hamiltonian are assumed to be $J = 4$ eV and $\xi = 0.05$ eV [52]. $\vec{L}$ and $\vec{S}$ are angular and spin momentum operators divided by $\hbar$ and thus unitless. In the exchange interaction term, a molecular field direction $\hat{\alpha} \in \{\hat{x}, \hat{y}, \hat{z}\}$ is given for the purpose of compelling the AFM easy axis of Fe spins. We use the basis as $|\Psi^{\alpha}\rangle = |d\rangle \otimes |s_{\alpha}\rangle$ where $|d\rangle \in \{|d_{x^2-y^2}\rangle, |d_{3z^2-r^2}\rangle, |d_{xy}\rangle, |d_{yz}\rangle, |d_{zx}\rangle\}$ represents the 3*d* real-orbital state suitable for an orbital in the crystal field of oxygen octahedron and $|s_{\alpha}\rangle \in \{|\uparrow_{\alpha}\rangle, |\downarrow_{\alpha}\rangle\}$ stands for the spin eigenstate when it is measured along the direction $\hat{\alpha}$. The Hamiltonian is represented in the basis and then diagonalized to obtain ten eigenvalues and eigenvectors. We obtain the total energy with occupying the lowest five and charge-transferred electronic states at the three different molecular field directions in order to compare them and determine the most preferred AFM spin axis. The detailed explanation of each Hamiltonian term is described below.

To calculate the crystal field potential, we assume that six $O^{2-}$ ions and eight $Bi^{3+}$ ions as point charges, and their positions are given in both the three dimensional Cartesian coordinate $(x, y, z)$ system and the spherical coordinate $(r, \theta, \phi)$ system where the origin $(0, 0, 0)$ is located at the $Fe^{3+}$ ion. Before describing the matrix elements of the crystal field potential within the $|d\rangle$ basis, we first compute the matrix elements of the crystal field potential under a basis $|n,l,m\rangle$ where $n$, $l$ and $m$ are the principal quantum number, the azimuthal quantum number and the magnetic quantum number, respectively. The Fe $3d$ levels correspond to $n = 3$, $l = 2$ and $m \in \{-2,-1,0,1,2\}$. The matrix elements can be expressed as [59]

$$V_{CF}(r,\theta,\phi) = \sum_i -\frac{Z_i e^2}{4\pi\varepsilon_0} \sum_{k=0}^{\infty} \frac{r^k}{R_i^{k+1}} \frac{4\pi}{2k+1} \sum_{m=-k}^{k} Y_{k,m}(\theta,\phi) Y_{k,m}^*(\theta_i,\phi_i), \tag{A2a}$$

$$\langle n,l,m'|V_{CF}|n,l,m\rangle = \iiint R_{n,l}^2(r,Z_{eff}) Y_{2,m'}^*(\theta,\phi) V_{CF}(r,\theta,\phi) Y_{2,m}(\theta,\phi) r^2 \sin\theta\, dr\, d\theta\, d\phi$$
$$= \sum_{k\in\{0,2,\cdots,2l\}} \gamma_{k,m'-m}(R) G_{k,2,m',2,m} U_{n,l,k}(R,Z_{eff}), \tag{A2b}$$

where $Z_i$ and $(R_i, \theta_i, \phi_i)$ are the charge and the spherical coordinate of $i^{th}$ point charges including $O^{2-}$ and $Bi^{3+}$ ions, and $\gamma_{k,m}$, $G_{k,l',m',l,m}$ and $U_{n,l,k}$ are written as

$$\gamma_{k,m}(R) = \sqrt{\frac{4\pi}{2k+1}} \sum_i Z_i \left(\frac{R}{R_i}\right)^{k+1} Y_{k,m}^*(\theta_i,\phi_i), \tag{A2c}$$

$$G_{k,l',m',l,m} = (-1)^{m'} \sqrt{(2l'+1)(2l+1)} \begin{pmatrix} l' & k & l \\ 0 & 0 & 0 \end{pmatrix} \begin{pmatrix} l' & k & l \\ -m' & m'-m & m \end{pmatrix}, \tag{A2d}$$

$$U_{n,l,k}(R,Z_{eff}) = -\frac{e^2}{4\pi\varepsilon_0 R^{k+1}} \int R_{n,l}^2(r,Z_{eff}) r^{k+2} dr. \tag{A2e}$$

In Eq. A2d, the Wigner 3-$j$ symbol is defined as

$$\begin{pmatrix} l_1 & l_2 & l_3 \\ m_1 & m_2 & m_3 \end{pmatrix} \equiv \frac{(-1)^{l_1-l_2-m_3}}{\sqrt{2l_3+1}} \langle l_3,-m_3|l_1,m_1;l_2,m_2\rangle, \tag{A2f}$$

where $\langle l_3,-m_3|l_1,m_1;l_2,m_2\rangle$ is the Clebsch-Gordan coefficient and $|l_1,m_1;l_2,m_2\rangle \equiv |l_1,m_1\rangle \otimes |l_2,m_2\rangle$ [60]. The Wigner 3-$j$ symbol vanishes unless $m_1 + m_2 + m_3 = 0$. Since Fe $3d$ levels correspond to the case where $n = 3$ and $l = 2$, only the $R_{3,2}(r,Z_{eff})$ term of the hydrogen-like radial wave functions in Eq. A2e is considered in our calculation:

$$R_{3,2}(r, Z_{eff}) = \frac{4}{81\sqrt{30}} \left( \frac{Z_{eff}}{a_0} \right)^{\frac{7}{2}} r^2 e^{-\frac{Z_{eff}}{3a_0} r}. \tag{A2g}$$

where $a_0$ is the Bohr radius. We set $Z_{eff} = 8$, i.e., the effective nuclear charge a Fe 3d electron experiences under the consideration that the $1s^2 2s^2 2p^6 3s^2 3p^6$ core electrons partially screen the nuclear charge of Fe ion. We note that the variation of the $Z_{eff}$ value within the range from 7 to 9 does not change our final conclusion on the most preferred AFM spin axis; the calculation with a lower $Z_{eff}$ value results in a larger single-ion anisotropy under the same Fe off-center displacement. Next, we transform the spherical harmonics basis set $\{|n,l,m\rangle\}$ to the real-orbital basis set $\{|d\rangle\}$ by using the relations below:

$$|d_{x^2-y^2}\rangle = \frac{1}{\sqrt{2}}(|n=3, l=2, m=2\rangle + |n=3, l=2, m=-2\rangle), \tag{A3a}$$

$$|d_{3z^2-r^2}\rangle = |n=3, l=2, m=0\rangle, \tag{A3b}$$

$$|d_{xy}\rangle = \frac{i}{\sqrt{2}}(|n=3, l=2, m=2\rangle - |n=3, l=2, m=-2\rangle), \tag{A3c}$$

$$|d_{yz}\rangle = \frac{i}{\sqrt{2}}(|n=3, l=2, m=1\rangle + |n=3, l=2, m=-1\rangle), \tag{A3d}$$

$$|d_{zx}\rangle = \frac{1}{\sqrt{2}}(|n=3, l=2, m=1\rangle - |n=3, l=2, m=-1\rangle). \tag{A3e}$$

The matrix form of the $-J\vec{S}\cdot\hat{\alpha}$ term in the Hamiltonian under the basis set $\{|\Psi^\alpha\rangle\}$ is simply written as

$$\sum_{\nu,\mu} |\Psi_\nu^\alpha\rangle \langle \Psi_\nu^\alpha | (-J\vec{S}\cdot\hat{\alpha}) | \Psi_\mu^\alpha \rangle \langle \Psi_\mu^\alpha | = \frac{1}{2} \begin{pmatrix} -J & 0 & 0 & 0 & 0 & 0 & 0 & 0 & 0 & 0 \\ 0 & -J & 0 & 0 & 0 & 0 & 0 & 0 & 0 & 0 \\ 0 & 0 & -J & 0 & 0 & 0 & 0 & 0 & 0 & 0 \\ 0 & 0 & 0 & -J & 0 & 0 & 0 & 0 & 0 & 0 \\ 0 & 0 & 0 & 0 & -J & 0 & 0 & 0 & 0 & 0 \\ 0 & 0 & 0 & 0 & 0 & J & 0 & 0 & 0 & 0 \\ 0 & 0 & 0 & 0 & 0 & 0 & J & 0 & 0 & 0 \\ 0 & 0 & 0 & 0 & 0 & 0 & 0 & J & 0 & 0 \\ 0 & 0 & 0 & 0 & 0 & 0 & 0 & 0 & J & 0 \\ 0 & 0 & 0 & 0 & 0 & 0 & 0 & 0 & 0 & J \end{pmatrix}, \tag{A4}$$

where the basis vectors are arranged in the following order: $|\Psi_1^\alpha\rangle = |d_{x^2-y^2} \uparrow_\alpha\rangle$, $|\Psi_2^\alpha\rangle = |d_{3z^2-r^2} \uparrow_\alpha\rangle$, $|\Psi_3^\alpha\rangle = |d_{xy} \uparrow_\alpha\rangle$, $|\Psi_4^\alpha\rangle = |d_{yz} \uparrow_\alpha\rangle$, $|\Psi_5^\alpha\rangle = |d_{zx} \uparrow_\alpha\rangle$, $|\Psi_6^\alpha\rangle = |d_{x^2-y^2} \downarrow_\alpha\rangle$, $|\Psi_7^\alpha\rangle = |d_{3z^2-r^2} \downarrow_\alpha\rangle$, $|\Psi_8^\alpha\rangle = |d_{xy} \downarrow_\alpha\rangle$, $|\Psi_9^\alpha\rangle = |d_{yz} \downarrow_\alpha\rangle$, and $|\Psi_{10}^\alpha\rangle = |d_{zx} \downarrow_\alpha\rangle$. The energy of the spin-up levels (where the spins are parallel to the molecular field $J\hat{\alpha}$) is shifted by $-J/2$ and that of spin-down levels (where the spins are antiparallel to the molecular field $J\hat{\alpha}$) is shifted by $J/2$ due to the term.

In order to get the matrix elements of the $\xi \vec{L} \cdot \vec{S}$ term in the Hamiltonian, we first expand it by using the Pauli matrices:

$$\sum_{s',s} |s'_z\rangle \langle s'_z | \xi \vec{L} \cdot \vec{S} | s_z\rangle \langle s_z| = \sum_{s',s} |s'_z\rangle \langle s'_z | \xi(L_x S_x + L_y S_y + L_z S_z) | s_z\rangle \langle s_z|$$

$$= \frac{\xi}{2}\left[ L_x \begin{pmatrix} 0 & 1 \\ 1 & 0 \end{pmatrix} + L_y \begin{pmatrix} 0 & -i \\ i & 0 \end{pmatrix} + L_z \begin{pmatrix} 1 & 0 \\ 0 & -1 \end{pmatrix} \right]. \quad (A5)$$

$$= \frac{\xi}{2}\begin{pmatrix} L_z & L_- \\ L_+ & -L_z \end{pmatrix}$$

From the above matrix, we realize that the spin-orbit coupling allows the mixing between spin-up and spin-down levels; in addition, the spin-orbit coupling term generates the non-zero angular momentum in the final eigenstates obtained by diagonalizing the Hamiltonian. Each matrix element of the spin-orbit coupling term within the basis set $\{|\Psi^z\rangle\}$ can be obtained by utilizing the above equation and the matrix is given by

$$\sum_{v,\mu} |\Psi_v^z\rangle \langle \Psi_v^z | \xi \vec{L} \cdot \vec{S} | \Psi_\mu^z\rangle \langle \Psi_\mu^z | =$$

$$\xi\begin{pmatrix}
0 & 0 & i & 0 & 0 & 0 & 0 & 0 & \frac{i}{2} & -\frac{1}{2} \\
0 & 0 & 0 & 0 & 0 & 0 & 0 & 0 & \frac{i\sqrt{3}}{2} & \frac{\sqrt{3}}{2} \\
-i & 0 & 0 & 0 & 0 & 0 & 0 & 0 & -\frac{1}{2} & -\frac{i}{2} \\
0 & 0 & 0 & 0 & -\frac{i}{2} & -\frac{i}{2} & -\frac{i\sqrt{3}}{2} & \frac{1}{2} & 0 & 0 \\
0 & 0 & 0 & \frac{i}{2} & 0 & \frac{1}{2} & -\frac{\sqrt{3}}{2} & \frac{i}{2} & 0 & 0 \\
0 & 0 & 0 & \frac{i}{2} & \frac{1}{2} & 0 & 0 & -i & 0 & 0 \\
0 & 0 & 0 & \frac{i\sqrt{3}}{2} & -\frac{\sqrt{3}}{2} & 0 & 0 & 0 & 0 & 0 \\
0 & 0 & 0 & \frac{1}{2} & -\frac{i}{2} & i & 0 & 0 & 0 & 0 \\
-\frac{i}{2} & -\frac{i\sqrt{3}}{2} & -\frac{1}{2} & 0 & 0 & 0 & 0 & 0 & 0 & \frac{i}{2} \\
-\frac{1}{2} & \frac{\sqrt{3}}{2} & \frac{i}{2} & 0 & 0 & 0 & 0 & 0 & -\frac{i}{2} & 0
\end{pmatrix}. \qquad (A6)$$

Since the matrix representation of the spin-orbit coupling term depends on the basis set we choose (however, the crystal field potential and the exchange interaction terms maintain the same matrix representation under any molecular field direction), we show the other two cases, *i.e.*, under the basis sets $\{|\Psi^x\rangle\}$ and $\{|\Psi^y\rangle\}$:

$$\sum_{\nu,\mu}|\Psi^x_\nu\rangle\langle\Psi^x_\nu|\xi\vec{L}\cdot\vec{S}|\Psi^x_\mu\rangle\langle\Psi^x_\mu|=$$

$$\xi\begin{pmatrix}
0 & 0 & 0 & \frac{i}{2} & 0 & 0 & 0 & i & 0 & \frac{1}{2} \\
0 & 0 & 0 & \frac{i\sqrt{3}}{2} & 0 & 0 & 0 & 0 & 0 & -\frac{\sqrt{3}}{2} \\
0 & 0 & 0 & 0 & -\frac{i}{2} & -i & 0 & 0 & \frac{1}{2} & 0 \\
-\frac{i}{2} & -\frac{i\sqrt{3}}{2} & 0 & 0 & 0 & 0 & 0 & -\frac{1}{2} & 0 & -\frac{i}{2} \\
0 & 0 & \frac{i}{2} & 0 & 0 & -\frac{1}{2} & \frac{\sqrt{3}}{2} & 0 & \frac{i}{2} & 0 \\
0 & 0 & i & 0 & -\frac{1}{2} & 0 & 0 & 0 & -\frac{i}{2} & 0 \\
0 & 0 & 0 & 0 & \frac{\sqrt{3}}{2} & 0 & 0 & 0 & -\frac{i\sqrt{3}}{2} & 0 \\
-i & 0 & 0 & -\frac{1}{2} & 0 & 0 & 0 & 0 & 0 & \frac{i}{2} \\
0 & 0 & \frac{1}{2} & 0 & -\frac{i}{2} & \frac{i}{2} & \frac{i\sqrt{3}}{2} & 0 & 0 & 0 \\
\frac{1}{2} & -\frac{\sqrt{3}}{2} & 0 & \frac{i}{2} & 0 & 0 & 0 & -\frac{i}{2} & 0 & 0
\end{pmatrix}, \qquad (A7)$$

$$\sum_{\nu,\mu}|\Psi^y_\nu\rangle\langle\Psi^y_\nu|\xi\vec{L}\cdot\vec{S}|\Psi^y_\mu\rangle\langle\Psi^y_\mu|=$$

$$\xi \begin{pmatrix} 0 & 0 & 0 & 0 & -\frac{i}{2} & 0 & 0 & i & \frac{1}{2} & 0 \\ 0 & 0 & 0 & 0 & \frac{i\sqrt{3}}{2} & 0 & 0 & 0 & \frac{\sqrt{3}}{2} & 0 \\ 0 & 0 & 0 & -\frac{i}{2} & 0 & -i & 0 & 0 & 0 & -\frac{1}{2} \\ 0 & 0 & \frac{i}{2} & 0 & 0 & -\frac{1}{2} & -\frac{\sqrt{3}}{2} & 0 & 0 & -\frac{i}{2} \\ \frac{i}{2} & -\frac{i\sqrt{3}}{2} & 0 & 0 & 0 & 0 & 0 & \frac{1}{2} & \frac{i}{2} & 0 \\ 0 & 0 & i & -\frac{1}{2} & 0 & 0 & 0 & 0 & 0 & \frac{i}{2} \\ 0 & 0 & 0 & -\frac{\sqrt{3}}{2} & 0 & 0 & 0 & 0 & 0 & -\frac{i\sqrt{3}}{2} \\ -i & 0 & 0 & 0 & \frac{1}{2} & 0 & 0 & 0 & \frac{i}{2} & 0 \\ \frac{1}{2} & \frac{\sqrt{3}}{2} & 0 & 0 & -\frac{i}{2} & 0 & 0 & -\frac{i}{2} & 0 & 0 \\ 0 & 0 & -\frac{1}{2} & \frac{i}{2} & 0 & -\frac{i}{2} & \frac{i\sqrt{3}}{2} & 0 & 0 & 0 \end{pmatrix}. \quad (A8)$$

We note that the above two spin-orbit coupling matrices are achieved by applying the spin-rotation relations to the $\hat{\alpha} = \hat{z}$ case:

$$|d \downarrow_x\rangle = \frac{1}{\sqrt{2}}(|d \uparrow_z\rangle - |d \downarrow_z\rangle), \quad (A9a)$$

$$|d \uparrow_x\rangle = \frac{1}{\sqrt{2}}(|d \uparrow_z\rangle + |d \downarrow_z\rangle), \quad (A9b)$$

$$|d \downarrow_y\rangle = \frac{1}{\sqrt{2}}(|d \uparrow_z\rangle - i|d \downarrow_z\rangle), \quad (A9c)$$

$$|d \uparrow_y\rangle = \frac{1}{\sqrt{2}}(|d \uparrow_z\rangle + i|d \downarrow_z\rangle). \quad (A9d)$$

By performing the numerical matrix calculation, the Hamiltonian is diagonalized and we are able to obtain a set of ten eigenvalues $\{E_1^\alpha, E_2^\alpha, \cdots, E_\nu^\alpha, \cdots, E_{10}^\alpha\}$ and the corresponding eigenvectors $|\Phi_\nu^\alpha\rangle = \sum_{\mu=1}^{10} c_{\nu,\mu}^\alpha |\Psi_\mu^\alpha\rangle$ (where $\nu \in \{1,2,\cdots,10\}$) for each molecular field direction. Then, as a starting point, we occupy electrons from the lowest eigenstate up to five eigenstates so that the five lowest eigenstates are fully occupied ($n_{\nu \in \{1,2,\cdots,5\}}^\alpha = 1$) because the Fe ion in BFO formally has five 3$d$ electrons in the limit of ionic picture.

Although our calculation in this ionic picture qualitatively gives the same result, we intend to pursue a more correct result by considering a covalent bonding effect. In addition to the five filled electronic states, we partially occupy the other five empty eigenstates in order to reflect the hybridization between empty Fe 3*d* orbitals and filled O 2*p* orbitals as expected from the previously reported O *K*-edge XAS spectra [43, 54]. It is assumed that the amount of electrons transferred from O 2*p* to Fe 3*d* is proportional to the square of the interatomic matrix elements. The interatomic matrix elements are described by the Slater-Koster two-center integrals [61] and scale with $R_i^{-7/2}$ according to the Harrison's method [62]. The single-ion anisotropy of a BFO unit cell with the Fe off-centering is estimated by comparing the total energy $E_{\text{tot}}^{\alpha} = \sum_{\nu=1}^{10} n_\nu^\alpha E_\nu^\alpha$ under three different molecular field directions ($\hat{\alpha} \in \{\hat{x}, \hat{y}, \hat{z}\}$).

**FIGURE CAPTIONS**

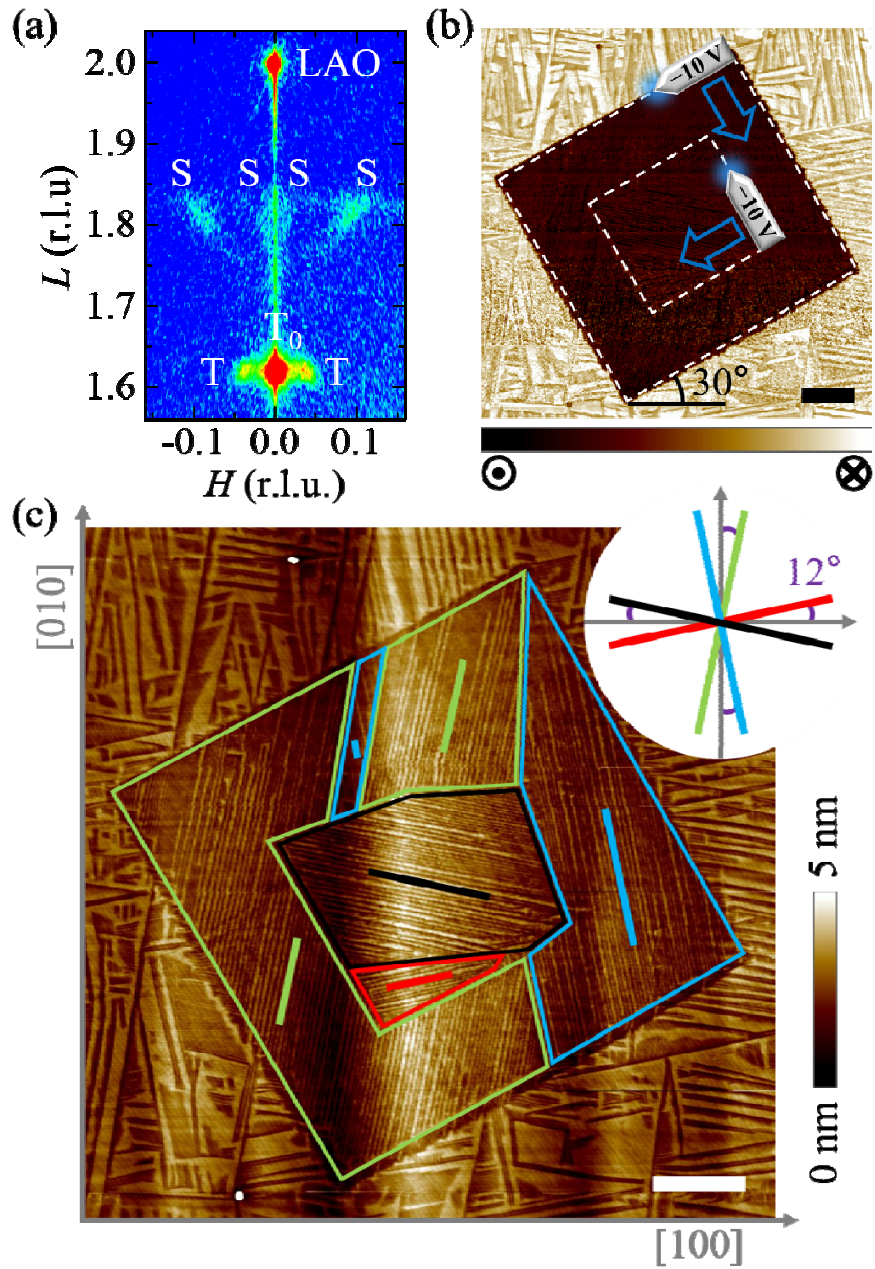

FIG. 1. (Color online) (a) Reciprocal space map around the (002) diffraction. The reciprocal lattice unit is defined as $2\pi/3.789$ Å$^{-1}$. (b) Out-of-plane PFM image of a double-box-poled region (white dashed boxes). The dc-bias voltages and two slow-scan axes during the double-box poling are provided. (c) Atomic force microscope image on the same region. Four well-aligned MPB regions are classified by four colored polygons and lines. Each MPB elongation axis has a 12°-offset from [100] or [010] crystalline axis. The scale bars indicate 2 μm.

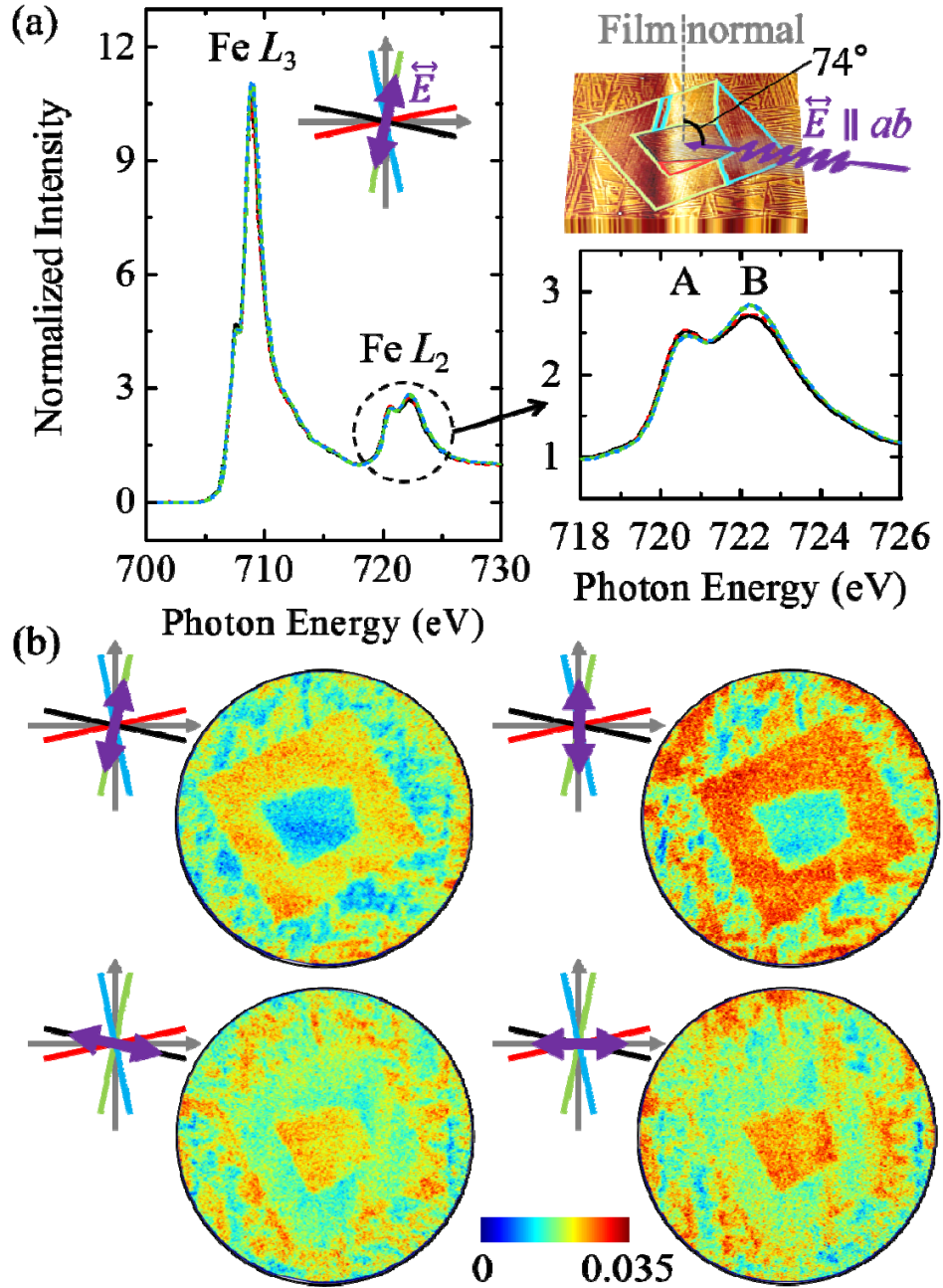

FIG. 2. (Color online) (a) Area-averaged Fe $L$-edge XAS spectra from the four electrically-aligned MBP regions with a $\sigma$-polarized x-ray beam where its electric-field axis is parallel to the green MPB elongation axis. On the right side, a zoomed-in graph is shown near the $L_2$ edge. (b) Four XLD-PEEM images at the double-box poled region. On the top-left of each image, a relation between an incident electric-field axis and the four MPB elongation axes is displayed. Here, the definition of the image contrast is $I_{\text{XLD-PEEM}} = (I_B - I_A)/(I_B + I_A)$.

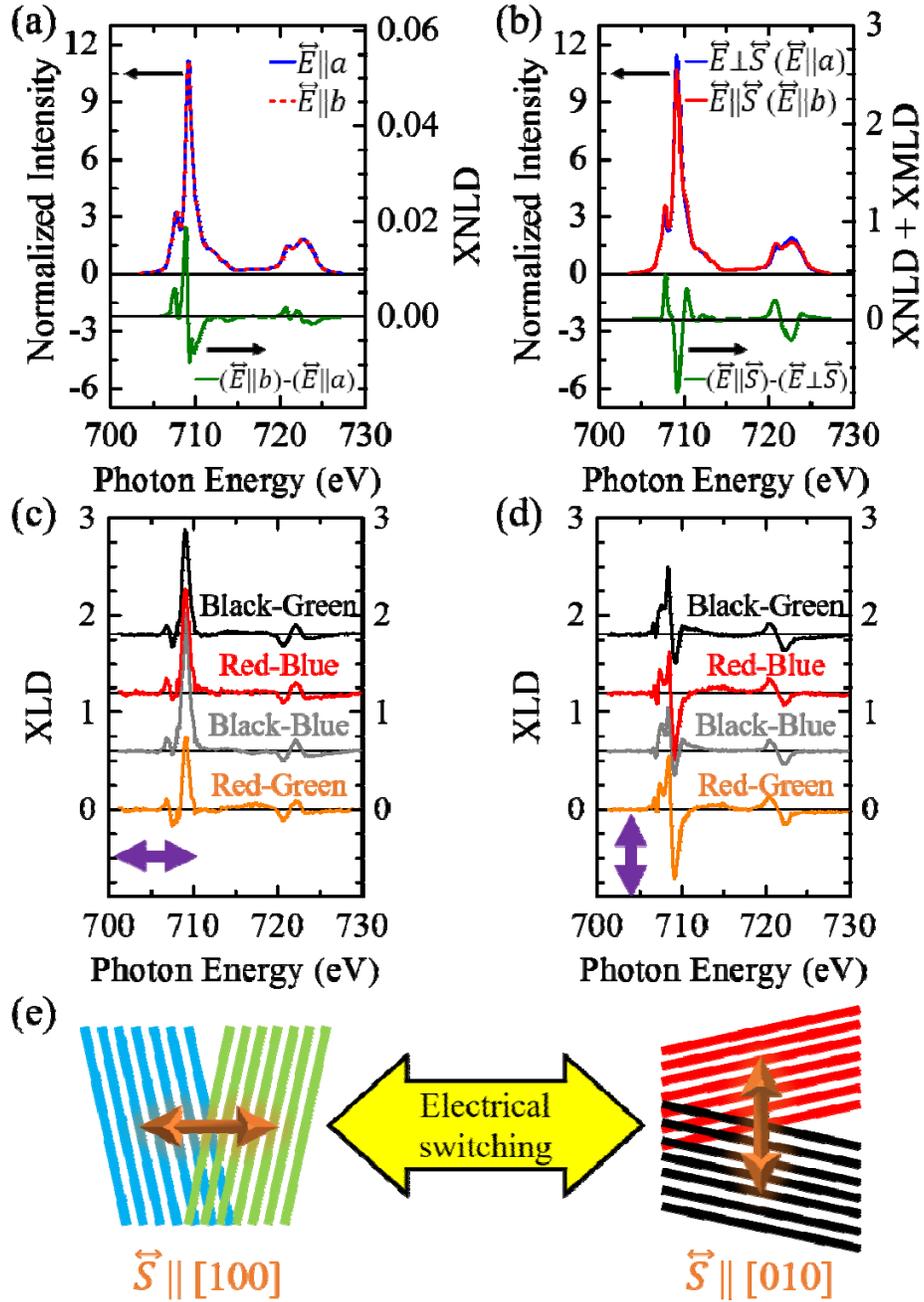

FIG. 3. (Color online) Cluster model calculations on a T-phase unit cell showing (a) XNLD and (b) XNLD + XMLD. (c), (d) Experimental XLD spectra when the electric field axis (purple-colored double arrow) of the incident x-ray beam is parallel to [100] or [010]. The experimental XLD signals are the difference between the XAS spectra of the two MPB regions. The color names (Black, Red, Green and Blue) correspond to the four MPB regions. The zero of vertical scale is set to the orange solid line and the other solid lines are shifted with a constant offset. (e) Schematic diagram showing the correlation between the MPB elongation axis and the AFM spin axis.

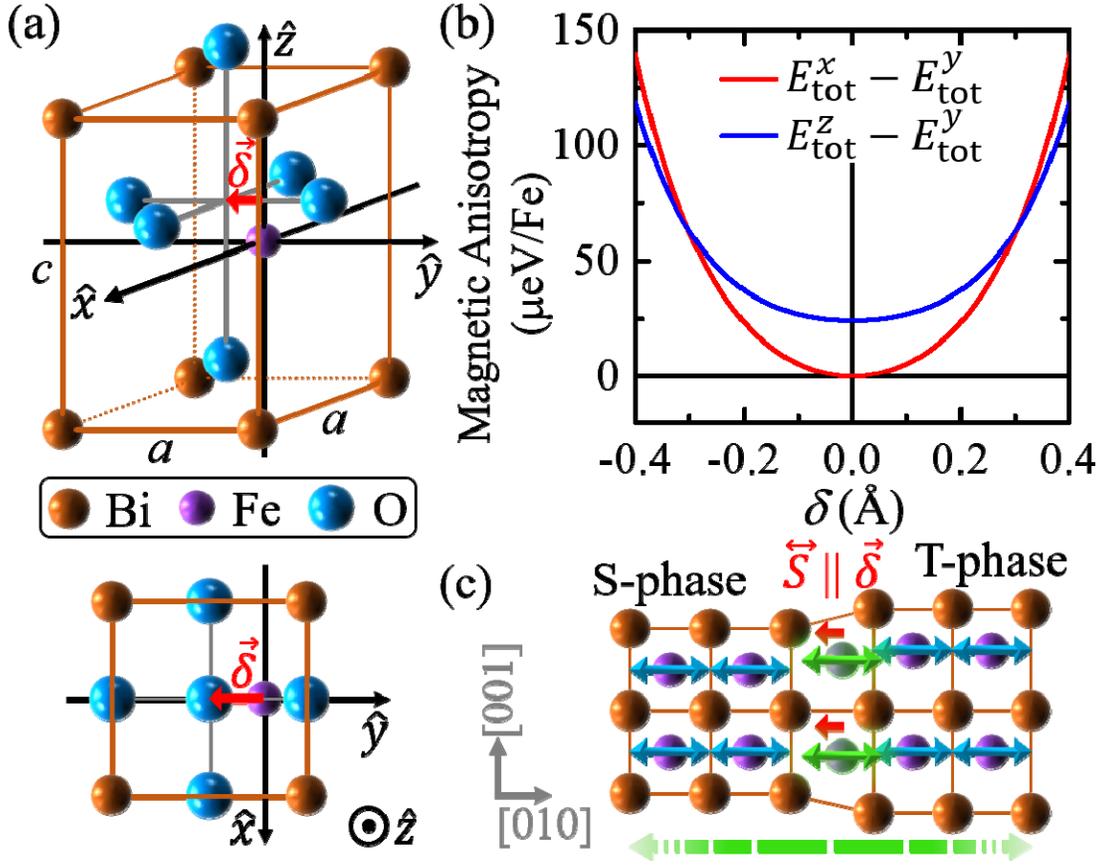

FIG. 4. (Color online) (a) Our structural model with a possible flexoelectric cation shift by $-\delta$ at MPB. (b) Calculated magnetic anisotropy as a function of $\delta$. (c) Schematic diagram of the magnetic easy axis at and near the MPB. Local AFM spin axis at MPB (expressed by green double-headed arrows) is compelled toward the axis of strain gradient due to the asymmetry in the environment of Fe ion and thereby inducing a non-quenched orbital angular momentum. Spins in the neighboring T and S-phase (blue double-headed arrows) are also aligned to the same axis through the intersite spin-spin interaction because the spins in the regions have less anisotropy within the *xy* plane at $\delta \sim 0$.

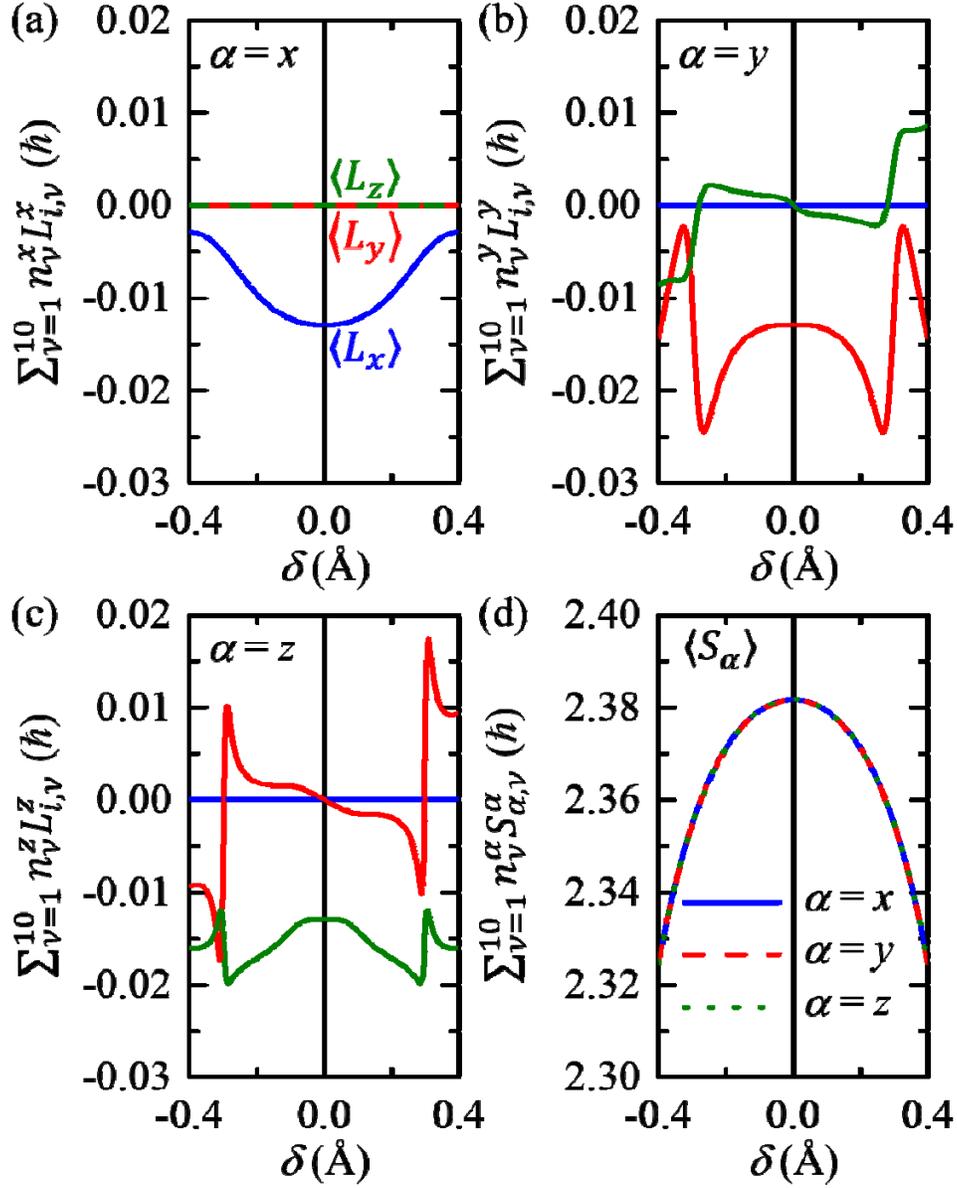

FIG. 5. (Color online) Three orbital angular momentum components as a function of the displacement $\delta$ along $y$-axis when spins are enforced to align along (a) $\hat{\alpha}=\hat{x}$, (b) $\hat{\alpha}=\hat{y}$, and (c) $\hat{\alpha}=\hat{z}$. The color label for each line is defined in (a). There exists the inversion of two energy levels between $d_{3z^2-r^2}$-orbital-dominant level and $d_{xy}$-orbital-dominant level at $\delta$ ~0.3 Å, making the evolution pattern of the orbital angular momentum components complex. (d) The spin component parallel to $\hat{\alpha}$ as a function of $\delta$ under the three different directions of $\hat{\alpha}$. The other spin components are almost suppressed due to the large molecular exchange field.